\documentclass[aps,prl,twocolumn,showpacs,superscriptaddress,preprintnumbers]{revtex4-1}
\usepackage{graphicx}
\usepackage{amsmath}
\usepackage[dvipsnames]{xcolor}
\usepackage{enumitem} 
\usepackage{hyperref}
\hypersetup{backref,colorlinks=true,urlcolor=blue,linkcolor=blue,citecolor=blue}
\newcommand{\be}{\begin{eqnarray}}
\newcommand{\ee}{\end{eqnarray}}

\begin{document}

\title{Two-flavor Simulations of the $\boldsymbol{\rho(770)}$ and the Role of
the $\boldsymbol{K\bar K}$ Channel}

\author{B. Hu}
\email{binhu@gwmail.gwu.edu}
\affiliation{
The George Washington University,
 Washington, DC 20052, USA}

\author{R. Molina}
\email{ramope71@email.gwu.edu}
\affiliation{
The George Washington University,
 Washington, DC 20052, USA}

\author{M.\ D\"oring}
\email{doring@gwu.edu}
\affiliation{
The George Washington University,
 Washington, DC 20052, USA}
\affiliation{Thomas Jefferson National Accelerator Facility, Newport News, VA
23606, USA}

\author{A. Alexandru}
\email{aalexan@gwu.edu}
\affiliation{
The George Washington University,
 Washington, DC 20052, USA}


\preprint{JLAB-THY-16-2316}

\begin{abstract}
The $\rho (770)$ meson is the most extensively studied resonance in lattice QCD
simulations in two ($N_f=2$) and three ($N_f=2+1$) flavor formulations. We
analyze $N_f=2$ lattice scattering data using unitarized Chiral Perturbation
Theory, allowing not only for the extrapolation in mass but also in flavor,
$N_f=2\to N_f=2+1$. The flavor extrapolation requires information from a global
fit to $\pi\pi$ and $\pi K$ phase shifts from experiment. While the chiral
extrapolation of $N_f=2$ lattice data leads to masses of the $\rho(770)$ meson
far below the experimental one, we find that the missing $K\bar{K}$ channel is
able to explain this discrepancy.
\end{abstract}


\maketitle


{\bf Introduction and Method.} With advances in algorithms and increasing
computational resources it has become feasible to extract phase shifts from 
lattice-QCD simulations.  The L\"uscher formalism relates the discrete energy
eigenvalues of the QCD Hamiltonian simulated in a finite box to phase shifts, 
up to contributions that are exponentially suppressed with the box
size~\cite{Luscher:1986pf, Luscher:1990ux}. Moving frames, twisted boundary
conditions or asymmetric boxes are means to extract more eigenvalues from the
same volume~\cite{Rummukainen:1995vs,  Liu:2005kr, Lage:2009zv, Bernard:2010fp,
Doring:2011vk, Doring:2011nd, Leskovec:2012gb, Doring:2012eu, Gockeler:2012yj}.
This allows to scan the amplitude at several energies over the resonance region,
which is a prerequisite for the reliable extraction of resonance masses and
widths.  An energy-dependent fit to extracted phase shifts is required, but it
is also possible to short-circuit the L\"uscher equation and fit amplitude
parameters directly to energy eigenvalues~\cite{Doring:2011vk}. 

More complicated multi-channel analyses have been carried out
recently~\cite{Dudek:2014qha, Wilson:2015dqa, Dudek:2016cru} that require in
most cases a parameterization in energy to relate different eigenvalues and thus
to compensate for missing information needed to describe such systems at a given
energy. Introducing an optical potential absorbs degrees of freedom without the
need of explicit parameterization, applicable to multi-channel but also
multi-particle systems~\cite{Agadjanov:2016mao, Lang:2014tia}. 

Yet, the simplest hadronic system containing a resonance, $I=1$ elastic $\pi\pi$
scattering via the $\rho(770)$ resonance, continues to be subject of recent
lattice QCD simulations providing a more and more accurate determination of the
amplitude and a test ground to benchmark new techniques.  Phase shifts for the
$I=1$ $\pi\pi$ interaction were extracted in calculations with two
mass-degenerate light  flavors ($N_f=2$)~\cite{Aoki:2007rd,Gockeler:2008kc,
Feng:2010es, Lang:2011mn, Pelissier:2012pi,  Bali:2015gji,Guo:2015dde, update}
and ones that include the strange quark flavor  ($N_f=2+1$)~\cite{Frison:2010ws,
Aoki:2011yj,Dudek:2012xn, Metivet:2014bga, Wilson:2015dqa,  Bulava:2016mks}. For
three flavors, the $\rho$ meson was analyzed and extrapolated using unitarized Chiral
Perturbation Theory (UCHPT)~\cite{Hanhart:2008mx, Nebreda:2010wv,
Pelaez:2010fj, Guo:2011pa, Guo:2012yt} in Refs.~\cite{Hanhart:2008mx, Nebreda:2010wv,
Pelaez:2010fj} and using CHPT with vector fields in
Refs.~\cite{Bruns:2013tja, Bruns:2004tj}. Finite-volume effects were studied in
Refs.~\cite{Albaladejo:2013bra, Chen:2012rp, Wu:2014vma}. The first
extrapolation of $N_f=2+1$ lattice phase shifts was recently performed in
Ref.~\cite{Bolton:2015psa}. See Refs.~\cite{Allton:2005fb, Armour:2005mk} for
the chiral extrapolation of partially quenched lattice results.

A recent lattice QCD study from the GWU group~\cite{update} noted that the
$\rho$ mass extracted from $N_f=2$ simulations is lighter than its physical
value. This is also supported by an independent calculation from the  RQCD
Collaboration~\cite{Bali:2015gji} very close to the physical mass, that finds an
even lighter mass for the $\rho$, albeit with larger error bars.

\setlength\belowcaptionskip{-3ex}
\setlength{\abovecaptionskip}{2pt}
\begin{figure}
\begin{center}
\includegraphics[width=0.99\linewidth]{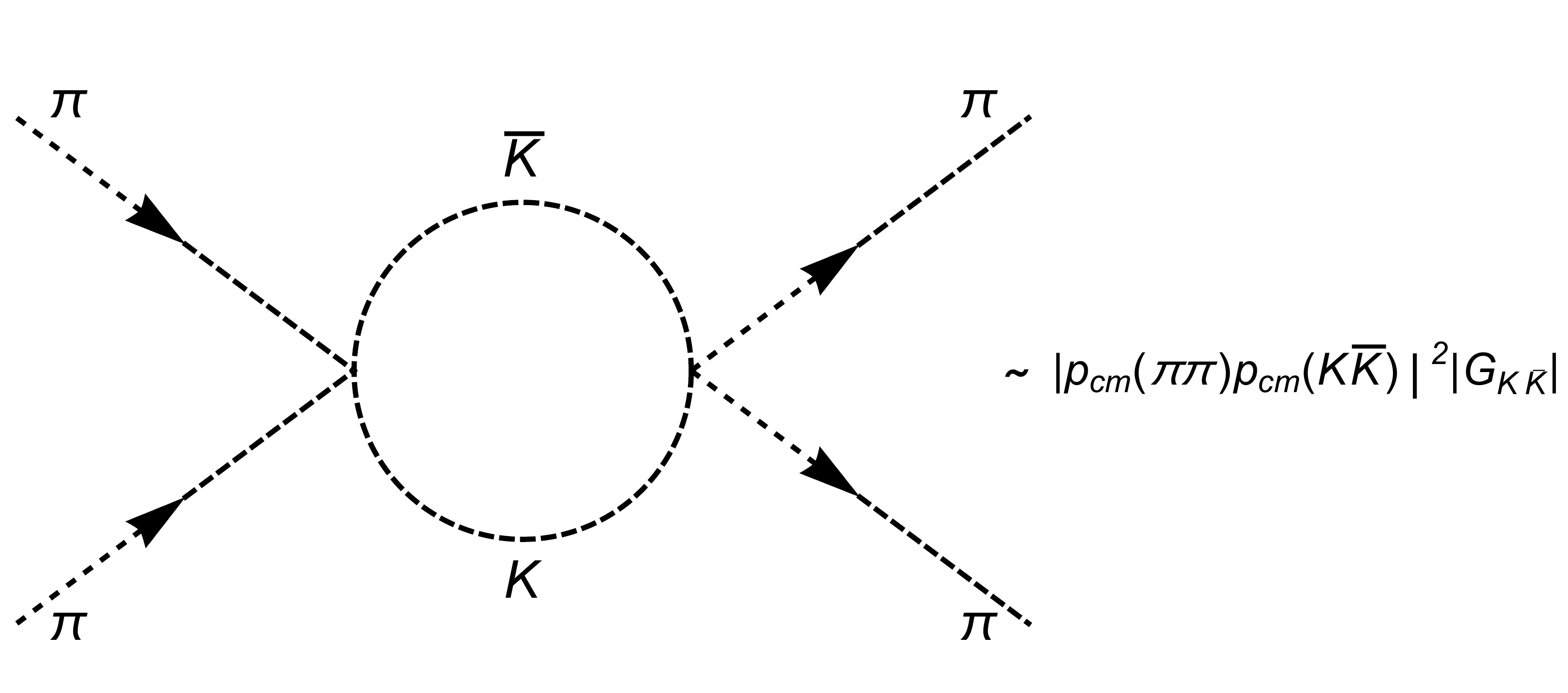}
\end{center}
\vspace*{-0.4cm}
\caption{Insertion of a $K\bar K$ intermediate state in $\pi\pi$
scattering, iterated in the present approach to provide coupled-channel unitarity.}
\label{fig:argument}
\end{figure}
In this letter, we discuss the hypothesis that the problem is tied to the
missing strange quark, or, formulated in terms of hadrons as degrees of freedom,
the absence of the $K\bar K$ channel. At first sight, the $K\bar K$ channel does
not seem to play a role; indeed, from the observed small inelasticity in the
$\rho$ channel~\cite{Protopopescu:1973sh} and the small $K\bar K$ phase shifts
obtained in analyses~\cite{Oller:1998hw, Doring:2013wka, Wilson:2015dqa}, it has
been conjectured that the $\rho$ effectively decouples from the $K\bar K$
channel. Yet, consider the insertion of an intermediate $K\bar K$ state in the
rescattering of two pions as displayed in Fig.~\ref{fig:argument}. The $p$-wave
dictates the behavior close to the thresholds according to $p_{\rm
cm}^2(\pi\pi)p_{\rm cm}^2(K\bar K)$ where $p_{\rm cm}$ are the momenta in the
center of mass. Additionally, there is a kaon loop $G_{K\bar K}$ including
dispersive parts~\cite{update}. The combined contribution exhibits a maximum
close to the $\rho$ mass. The full, unknown
interaction differs, of course, from this expression, but by a function that
varies only slowly with energy. In other words, while the effects from real
kaons are suppressed through the centrifugal barrier, virtual kaons can indeed
contribute to the $\pi\pi$ amplitude at the $\rho$ position, effectively
shifting its mass. However, a substantial shift of the $\rho$ mass might induce
a significant effect of the $K\bar K$ channel above threshold. Therefore, results
have to be checked with available constraints on $K\bar K$ phase shifts and inelasticities
from experiment and $N_f=2+1$ lattice QCD simulations (see discussions below and in 
supplemental material~\cite{supplement}).
 
In the present work, we use UCHPT to extrapolate $N_f=2$ lattice results to the
physical pion mass, but also to three flavors to study the role of the strange
quark. Phase shifts from $N_f=2+1$ simulations allow for direct extrapolation to
the physical point and will be subject of future studies. The discussed
mechanism of virtual or real intermediate $K\bar K$ states is taken into account
by choosing an SU(3) formulation in which $K\bar K$ states are re-summed to all
orders through unitarity in coupled channels. In the inverse amplitude
formulation and taking into account NLO contact interactions, this model has
been formulated in Ref.~\cite{Oller:1998hw}. We replace here
the cut-off with dimensional regularization~\cite{update}.
Also, there is a minor modification in $I=0$ $\pi K,\,\pi\eta$ scattering~\cite{Doring:2013wka}
at high energies.
For the extrapolation of $f_\pi$, the $M_\pi$ dependence of
Refs.~\cite{Gasser:1984gg, Nebreda:2010wv}, summarized in
Ref.~\cite{Molina:2015uqp}, is used. The workflow to extra\-polate to the
physical point is as follows (see also App.~B of Ref.~\cite{update} for a
similar procedure):

{\bf 1.} To fit the lattice phase shifts, the $K\bar K$ channel is removed from
the coupled-channel $\pi\pi/K\bar K$ system. The low-energy constants (LECs in
the definition of Ref.~\cite{update}) appear in two distinct linear combinations
in the $I=L=1$ $\pi\pi\to\pi\pi$ transition, $\hat l_1=2L_4+L_5$, $\hat
l_2=2L_1-L_2+L_3$,
which are the two fit parameters used (not to be confused with SU(2) CHPT LECs).

{\bf 2.}
$N_f=2$ lattice phase shifts are fitted including the known correlation  between
energy and phase shift (the error bars in the $(W,\delta (W))$ plane are
effectively inclined); in case of Refs.~\cite{update, Bali:2015gji}, the
covariance of energy eigenvalues is included in the fit. Data to be included in
the fit are chosen in the maximal range around the resonance position, in which
the fit passes Pearson's $\chi^2$ test at a 90\% upper confidence limit.

{\bf 3.}
The result is extrapolated to the physical pion mass $M_\pi=138$~MeV and then
the channel transitions $\pi\pi\to K\bar K$ and $K\bar K\to K\bar K$ are
switched on.  The combinations of LECs appearing in these transitions are
different from those of $\pi\pi\to\pi\pi$ and taken from a global fit to
experimental $\pi\pi$ and $\pi K$ phase shift data similar to that of
Ref.~\cite{Doring:2013wka}. 

{\bf 4.}
The solution is given by LECs and their uncertainties at the physical point. To
translate the results to the commonly used notation, all phase shifts are fitted
with the usual Breit-Wigner (BW) parameterization in terms of $g$ and $m_\rho$
(see, e.g., Ref.~\cite{update}) although they could be quoted in terms of pole
positions and residues which becomes increasingly popular~\cite{PDG,
Bolton:2015psa, Ronchen:2014cna}.


\bigskip

{\bf Results.} Available $N_f=2$ phase shift data~\cite{Aoki:2007rd,
Feng:2010es, Lang:2011mn,  Bali:2015gji, update} are analyzed. The data of
Ref.~\cite{Pelissier:2012pi} are not considered since they are superseded by
those of  Ref.~\cite{update}.  Results of Ref.~\cite{Feng:2010es} for larger
pion masses are not analyzed, because they are beyond the expected applicability
of the chiral extrapolation. 

\begin{figure}
\begin{center}
\includegraphics[width=1.\linewidth]{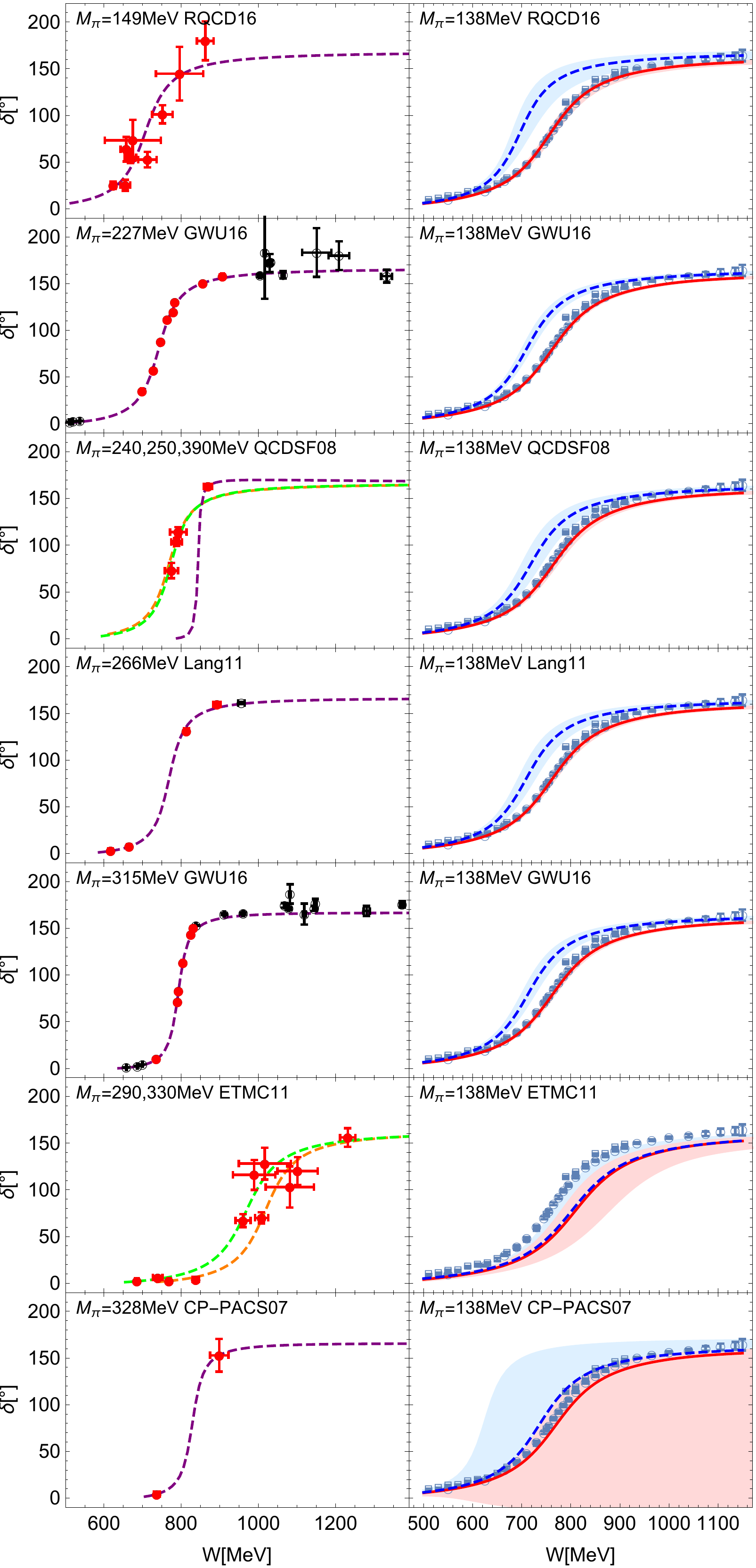}
\end{center}
\caption{Results 
for the $N_f=2$ lattice simulations (ordered by pion mass) of  Bali {\it et
al.}/RQCD~\cite{Bali:2015gji},  Guo {\it et al.}/GWU~\cite{update},  G\"ockeler
{\it et al.}/QCDSF~\cite{Gockeler:2008kc}, Lang {\it et
al.}~\cite{Lang:2011mn},  Feng {\it et al.}/ETMC~\cite{Feng:2010es},  Aoki {\it
et al.}/CP-PACS~\cite{Aoki:2007rd}. For each result, the left picture shows the
lattice data and fit, the right figure shows the $N_f=2$ chiral extrapolation
(blue dashed line/light blue area). Without changing this result, the $K\bar K$
channel is then included to predict the effect from the missing strange quark
(red solid line/light red area).  Experimental data (blue circles
from~\cite{Estabrooks:1974vu}, squares from  \cite{Protopopescu:1973sh}) are
then post-dicted. For inherent model uncertainties, see text. 
}
\label{fig:results1}
\end{figure}

The extrapolations for the different $N_f=2$ simulations are shown in
Fig.~\ref{fig:results1}. For each simulation, the left picture shows the lattice
phase shifts and fit (only best fit shown). Phases included in the fit,
according to the discussed criterion, are highlighted.  As for consistency of
the performed fits, the 68\% confidence ellipses in $\hat l_1,\hat l_2$ from
RQCD~\cite{Bali:2015gji}, GWU~\cite{update} ($m_\pi=227$~MeV and
$m_\pi=315$~MeV), Lang et al.~\cite{Lang:2011mn}, and CP-PACS~\cite{Aoki:2007rd}
all have a common overlap; the ellipse from QCDSF~\cite{Gockeler:2008kc} is very
slightly off, while the one from ETMC~\cite{Feng:2010es} is clearly incompatible
(see supplemental material for a picture~\cite{supplement}).

The right pictures of Fig.~\ref{fig:results1} show the $N_f=2$ chiral
extrapolation to the physical mass (blue dashed line/light blue area). Then,
without changing that extrapolated result, the UCHPT prediction for the missing
strange quark is included in terms of the $K\bar K$ channel (solid red line).
Experimental data~\cite{Estabrooks:1974vu, Protopopescu:1973sh} are post-dicted.

In the following discussion of results, we exclude the data from Aoki {\it et
al.}/PACS-CS~\cite{Aoki:2007rd} (2 measured phase shifts fitted with 2
parameters) and Feng {\it et al.}/ETMC~\cite{Feng:2010es} because the
uncertainties are very large, even when simultaneously fitting data from two
different pion masses.  As Fig.~\ref{fig:results1} reveals, the $N_f=2$
extrapolations all lead to a $\rho$ mass lighter than experiment. This is
particularly clear for the results of 
Refs.~\cite{Bali:2015gji,update,Lang:2011mn} where the computed $\rho$ mass is
lighter than the experimental one even before extrapolating to physical quark
mass, so that the extrapolation cannot be responsible for this discrepancy.

Switching on the $K\bar K$ channel shows significant effects and increases the
$\rho$ mass, leading in all but the excluded cases to a much improved
post-diction of the experimental data.  For the lattice data by Bali {\it et
al.}/RQCD~\cite{Bali:2015gji}, taking the covariances of energy eigenvalues into
account narrows the band of uncertainties by about 30\%. The phase shifts of
Ref.~\cite{update} are the most accurate ones, leading to a narrow band in the
final result. In fact, the lattice data uncertainties are so small that beyond
the region of $\pm 2\Gamma$ around the resonance mass, the fit does not pass the
mentioned $\chi^2$ test and therefore data are not included. In any case, we
have also fitted all phase shifts and found that the best fits for the two pion
masses barely change.  As the fits of the two pion masses of Ref.~\cite{update}
are consistent, a common fit has been carried out in Ref.~\cite{update} leading
to the most constraining results on the $\hat l_i$ and the chiral and flavor
extrapolations. As for the simulation by G\"ockeler {\it et
al.}/QCDSF~\cite{Gockeler:2008kc} we have also included the data from
$M_\pi=390$~MeV in a combined fit, which significantly reduces uncertainties.
The best fit barely changes when fitting only the $M_\pi=240,\,250$~MeV phases.
The highest data point by Lang {\it et al.}~\cite{Lang:2011mn} needed to be
removed to fulfill the mentioned $\chi^2$ test. Including this point barely
changes the central result (red line) but only leads to smaller uncertainties.
The solution is also stable when removing the second-highest point and keep the
highest.  

For all fits, we have also checked  that the inelasticity from the $K \bar K$
channel does not become larger than the observed total
inelasticity~\cite{Protopopescu:1973sh} up to $W\sim 1.15$~GeV. The $\omega\pi$
contribution to the latter has been evaluated in Ref.~\cite{Niecknig:2012sj}
(see also Ref.~\cite{Bruns:2013tja}). Our $K\bar K$ inelasticity is rather of
similar size as the $K\bar K$ inelasticity derived in
Ref.~\cite{Niecknig:2012sj} from the Roy-Steiner solution of
Ref.~\cite{Buettiker:2003pp}. The inelasticity is in any case smaller than the
bound quoted in Ref.~\cite{Eidelman:2003uh}. Yet, $4\pi$ channels are omitted in
the current work because the fitted lattice phase shifts are situated below
finite-volume thresholds, except for the highest energy of
Ref.~\cite{Bali:2015gji} (omitting this point does not change the best fit). 
The $4\pi$ channels are effectively absorbed in the LECs in the lattice fits,
but introduce some uncertainty in the chiral extrapolation. 

In the supplemental material~\cite{supplement}, the inelasticity is shown with
experiment~\cite{Protopopescu:1973sh} and also with the $N_f=2+1$ lattice
simulation of Ref.~\cite{Wilson:2015dqa} at $M_\pi=236$~MeV. Inelasticities are
well predicted and the small $K\bar K$ phase shift has even the same size and
sign as in Ref.~\cite{Wilson:2015dqa}.

The predicted $\pi\pi$ scattering lengths are close to the ${\cal O}(p^4)$ CHPT
value but some are just outside the 1-$\sigma$ range of the experimental result,
while effective ranges are of similar size as the ${\cal O}(p^6)$ CHPT
value~\cite{Bijnens:1997vq} as quoted in the supplemental
material~\cite{supplement}.

\setlength\belowcaptionskip{-3ex}
\setlength{\abovecaptionskip}{-7pt}
\begin{figure}
\begin{center}
\includegraphics[width=1.\linewidth]{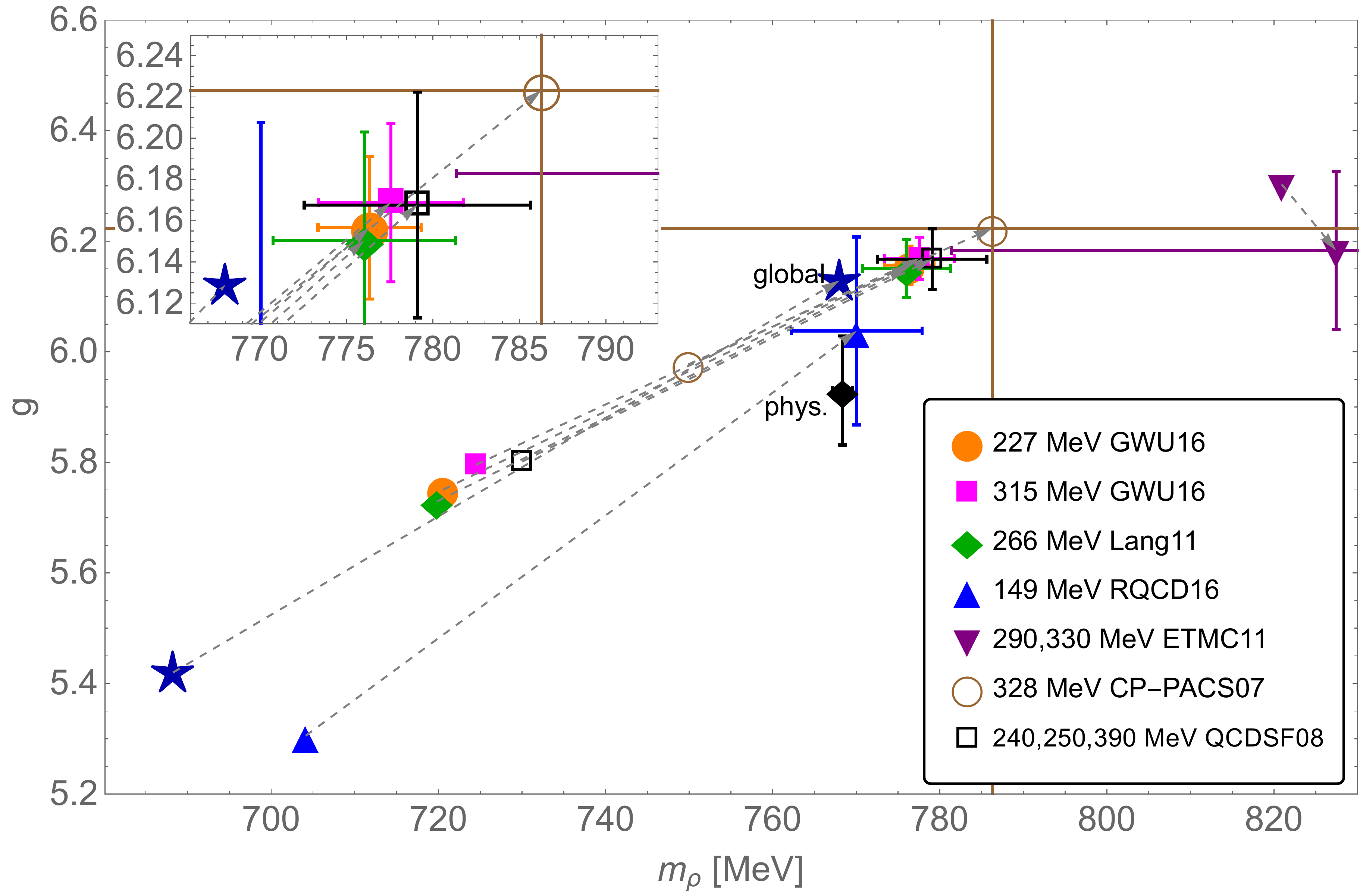}
\end{center}
\caption{Effect of the $K\bar K$ channel in the $(m_\rho,g)$ plane indicated
with arrows, after chiral extrapolation to the physical pion mass.  See
Fig.~\ref{fig:results1} for the labeling of the extrapolations. Only statistical
uncertainties are shown, and only for the case after including $K\bar K$. See
text for further explanations and supplemental material~\cite{supplement} for
the effect of the chiral extrapolation.
}
\label{fig:results2}
\end{figure}
In Fig.~\ref{fig:results2} we show the effect of the $K\bar K$ channel in the
$(m_\rho,g)$ plane. Remember that $(m_\rho,g)$ emerge from Breit-Wigner fits to
the UCHPT solutions. This is also the case for the experimental point, indicated
as ``phys.''. The comparability of all shown $(m_\rho,g)$ with other values in
the literature is therefore limited but in practice quite accurate.

To keep the figure simple, no error bars are shown for the chirally extrapolated
results; see previous remark on consistency of the fits. Once the $K\bar K$
channel is switched on, Fig.~\ref{fig:results2} shows that $g$ and $m_\rho$ are
slightly over-extrapolated. A possible reason is model deficiency. On one hand,
problems could originate from the formulation: we include NLO contact
terms~\cite{Oller:1998hw} but not the one-loop contributions at NLO as in
Ref.~\cite{Nebreda:2010wv}. On the other hand, the LECs entering the $\pi\pi\to
K\bar K$ and $K\bar K\to K\bar K$ transitions are not fully determined from the
fit of $N_f=2$ lattice data and therefore taken from a global fit to
experimental $\pi\pi$ and $\pi K$ phase shifts in different isospin and angular
momentum, similar to that of Ref.~\cite{Doring:2013wka}.  That global fit
compromises between different data sets, leading to a slightly wider $\rho$
resonance. In Fig.~\ref{fig:results2}, a Breit-Wigner fit to that solution is
indicated as ``global'' with a star. Indeed, the value for $g$ is slightly too
large. In any case, it is instructive to remove here the $K\bar K$ channel. As
Fig.~\ref{fig:results2} shows (star at $m_\rho\approx 690$~MeV), the result
(again, deduced only from experimental information) exhibits the same trend as
the $N_f=2$ lattice data, i.e., a lighter and narrower $\rho$.

Inherent model uncertainties from the $2\to 2+1$ flavor extrapolation can be
roughly estimated as in Ref.~\cite{update} by inserting the fitted $\hat l_1$,
$\hat l_2$ in the $\pi\pi\to K\bar K$ and $K\bar K\to K\bar K$ transitions,
instead of taking them from the global fit to experimental data. As a result,
instead of over-extrapolating in $m_\rho$ and $g$, these quantities are now
mostly under-extrapolated. The observed differences translate into 
model/systematic uncertainties of comparable size as the statistical
uncertainties shown in Fig.~\ref{fig:results2} (see supplemental material for
values~\cite{supplement}). 

As part of the $\rho$ mass shift originates from the regularized $K\bar K$ 
propagator~\cite{update}, we also test the dependence of the results on the
value of the subtraction constant, changing it from the default value
$a=-1.28$~\cite{update} to $a=-0.8$ and $a=-1.7$. The global fits to
experimental phase shifts visibly deteriorate for these extreme values, e.g.,
for $\pi K$ scattering, but barely change in the $\rho$ channel as experimental
phase-shift data are more precise.  Following the described workflow, we find
changes of the final results of less than 10 MeV in $m_\rho$ and less than
$0.08$ in $g$.

In conclusion, the present results demonstrate the relevance of the $K\bar K$
channel, that can explain the systematically small lattice $\rho$ masses at the
physical point after the chiral SU(2) extrapolation. From the discussion, it
becomes clear that a full one-loop calculation~\cite{GomezNicola:2001as, 
Nebreda:2010wv} for confirmation and further improvement of the present results 
is desirable. A rough estimate for neglected changes in the $\pi\pi\to\pi\pi$ 
transition when including the strange quark can be obtained by using the 
SU(2)-SU(3) matching relations for LECs~\cite{Gasser:1984gg}, 
resulting in very small changes, of less than 1 MeV, in the $\rho$ masses.


\bigskip
\bigskip

{\bf Summary:} All accessible phase shift data on the $\rho$ meson from $N_f=2$
lattice QCD simulations  are analyzed using the inverse amplitude method
including NLO terms from Chiral Perturbation Theory. The $N_f=2$ fits are
extrapolated to the physical pion mass, and the $K\bar K$ channel is
subsequently switched on without further changing the fit parameters. For this
step, combinations of SU(3) low-energy constants, that are not accessible
through the $N_f=2$ lattice data, are taken from a global fit to experimental
meson-meson phase shifts. The $K\bar K$ channel improves the extrapolations of
the $\rho$ mass significantly except when the lattice data have large
uncertainties.


\begin{acknowledgments}

This work is supported by the National Science Foundation (CAREER grants 
PHY-1452055 and PHY-1151648, PIF grant No. 1415459) and by GWU (startup grant).
M.D. is also supported by the U.S. Department of Energy, Office of Science, 
Office of Nuclear Physics under contract DE-AC05-06OR23177. 
A.A. is supported in part by  the U.S. Department of Energy grant DE-FG02-95ER40907.
The authors thank G. Bali and A. Cox for providing jackknife ensembles to
include correlations of energy eigenvalues, and G. Bali, D. Guo, Z.-H. Guo, C.
Hanhart, B. Kubis, C. Lang, M. Mai, U.-G. Mei{\ss}ner, D. Mohler, E. Oset, S.
Prelovsek, J. Ruiz de Elvira, and A. Rusetsky for discussions.

\end{acknowledgments}

\vspace*{-0.5cm}


\end{document}